# Revisited storage and dynamic recovery of dislocation density evolution law : toward a generalized Kocks–Mecking model of strain-hardening

O. Bouaziz[*1,2]

[1]*ArcelorMittal Research, Voie Romaine-BP30320, 57283 Maizières-lès-Metz Cedex, France*
[2]*Centre des Matériaux, Ecole des Mines de Paris, CNRS UMR 7633, B.P. 87, 91003 Evry Cedex, France*

**Abstract**
The current work extends the well established approach of Kocks and Mecking by a more realistic description of strain-hardening using an original dislocation density law with a revisited physical understanding of dynamic recovery, without new material parameters and keeping only one internal variable.

**Introduction**

The understanding of strain-hardening of metals is one of main important field of research in physical metallurgy. The Kocks-Mecking (KM) approach [1-2] is now the main way to have a physical based description of the strain-hardening due to the storage of dislocations induced by plastic strain. In the case of coarse grain polycristal, the evolution of the average dislocation density is expressed as the competition between a storage term and an annihilation term :

$$\frac{d\rho}{d\varepsilon} = M\left(\frac{d\rho^+}{d\varepsilon} - \frac{d\rho^-}{d\varepsilon}\right) \quad (1)$$

expressed as :

$$\frac{d\rho}{d\varepsilon} = M\left(\frac{k}{b}\sqrt{\rho} - f\rho\right) \quad (2)$$

where M is the Taylor factor, k is a constant describing the dislocation accumulation due to the interaction with forest dislocations as obstacles, b the Burgers vector and f is a parameter describing the dislocation annihilation due to dynamic recovery. In a second step the KM approach uses the Taylor equation linking the flow stress to the dislocation density:

$$\sigma = \alpha.M.\mu.b.\sqrt{\rho} \quad (3)$$

with $\alpha$ being a parameter that describes the interactions between the forest dislocations.
From Eq3, the strain-hardening is expressed as:

$$\frac{d\sigma}{d\varepsilon} = \frac{M.\alpha.\mu.b.}{2\sqrt{\rho}} \cdot \frac{d\rho}{d\varepsilon} \quad (4)$$

Finally the strain hardening can be also expressed as a function of flow stress :

$$\frac{d\sigma}{d\varepsilon} = \theta_o \cdot \left(1 - \frac{\sigma}{\sigma_s}\right) \tag{5}$$

where $\theta_o$ and $\sigma_s$ are two parameter function of k and f, $\theta_o$ being the strain-hardening in stage II.

Experimentally it has been observed that the linear decrease in the strain-hardening with the flow stress is only observed for low plastic strain (usually lower than 15%). After this range of deformation Eq5 understimates the strain-hardening. This is the reason why a new dislocation density evolution has to be proposed improving the predictions but keeping the simplicity of the original K-M approach : not more than 2 parameters and a clear physical meaning.

It is now considered the new dislocation evolution law expressed as :

$$\frac{d\rho}{d\varepsilon} = M\left(\frac{k}{b}\sqrt{\rho} \cdot \exp\left(-\xi \cdot \sqrt{\rho}\right)\right) \tag{6}$$

where k has the same meaning than in Eq2 and $\xi$ a characteristic length scale.

As $0 \leq \exp(-\xi \cdot \sqrt{\rho}) \leq 1$, the physical interpretation :

$$\frac{d\rho}{d\varepsilon} = M \cdot \left(\frac{d\rho^+}{d\varepsilon} - (1-F) \cdot \frac{d\rho^+}{d\varepsilon}\right) \tag{7}$$

With 1-F the fraction of stored dislocations annihilated by dynamic recovery.

As the mean distance between dislocations is $\frac{1}{\sqrt{\rho}}$, the function F can be seen as a Poissonian probability to annihilate a length of dislocation $\frac{1}{\sqrt{\rho}}$ in a region of diameter $\xi$. This the reason why $\xi$ is the capture distance for dynamic recovery.

It is interesting to notice that if $1 \gg \xi \cdot \sqrt{\rho}$, a first order expansion gives :

$$\frac{d\rho}{d\varepsilon} = M\left(\frac{k}{b}\sqrt{\rho} \cdot \left(1 - \xi \cdot \sqrt{\rho}\right)\right) \tag{8}$$

or

$$\frac{d\rho}{d\varepsilon} = M\left(\frac{k}{b}\sqrt{\rho} - \frac{k}{b} \cdot \xi \cdot \rho\right) \tag{9}$$

Which is asymptotycally the original Kocks-Mecking evolution law.

By integration of Eq.6, the strain-hardening can be expressed as:

$$\frac{d\sigma}{d\varepsilon} = \theta_o \cdot \exp\left(-\frac{\sigma}{\sigma_s}\right) \tag{10}$$

where $\theta_o$ and $\sigma_s$ are two parameter function of k and $\xi$.

Finally the behaviour law is :

$$\sigma = \sigma_s . \ln\left(1 + \frac{\theta_o}{\sigma_s}.\varepsilon\right) \quad (11)$$

In Fig1, taking the same values for $\theta_o$ and $\sigma_s$, the evolution of strain-hardening as a function of stress have been compared using Eq5 for Kocks-Mecking approach and Eq10 for the new one. It is highlighted that the two approaches gives similar results for low flow stress but the new one avoid a too rapid saturation of strain-hardening for higher stresses which is more consistent with all the available experimental data.

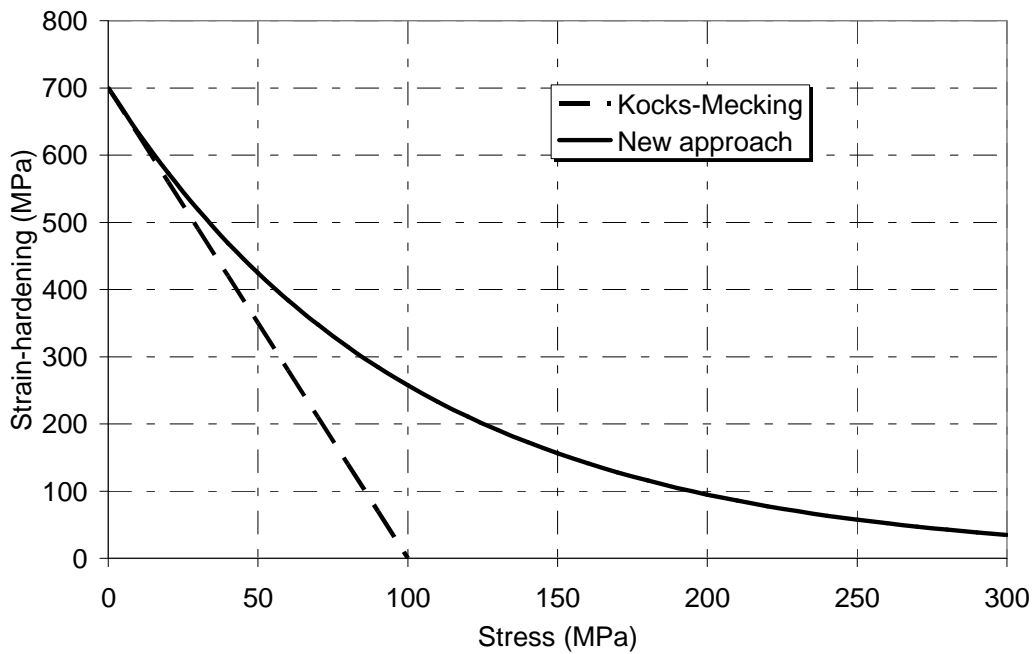

Figure 1 : Comparison between Kocks-Mecking and the new approach with same values for parameters ($\theta_o = 700$MPa and $\sigma_s = 100$).

**Conclusions**

By revisiting the process of dynamic recovery, an extension of the well established approach of Kocks and Mecking for the prediction of strain-hardening using dislocation density as internal variable has been proposed. Without new material parameters and keeping only one internal variable, an original dislocation density law has been assessed providing an obvious improvement of the modelling. As the new approach is easy-use, its applications to different metals and alloys present no problem.